\documentclass[prl,twocolumn,nopacs,preprintnumbers,notitlepage,amsmath,amssymb,superscriptaddress,floatfix]{revtex4-2}

\usepackage[utf8]{inputenc}
\usepackage{graphicx}
\usepackage{xcolor}
\usepackage[final]{changes} 

\newcommand{\add}[1]{\added{#1}}
\newcommand{\del}[1]{\deleted{#1}}

\makeatletter
\newcommand{\SItableofcontents}{%
  \section*{Contents}%
  \@starttoc{sitoc}%
}
\makeatother

\begin{document}

\title{Directed Cascades Generate New Critical Universality Classes}

\author{Didier Sornette}
\affiliation{Institute of Risk Analysis, Prediction and Management (Risks-X), Academy for Advanced Interdisciplinary Studies, Southern University of Science and Technology, Shenzhen, China}
\author{Eugenio Lippiello}
\affiliation{Department of Mathematics and Physics, University of Campania ``Luigi Vanvitelli'', Italy}
\author{Giuseppe Petrillo}
\affiliation{CNRS, ENS de Lyon, LPENSL, UMR 5672, 69342 Lyon, France}

\date{\today}

\begin{abstract}
Spectrally stable cascades can undergo enormous transient growth before eventually dying out. We show that this amplification is controlled by the directed architecture of the interactions, independently of the eigenvalues that determine asymptotic stability. More remarkably, when several critical subsystems are connected sequentially, their depth becomes a new control parameter for critical fluctuations: each additional critical stage generates a new cascade-size exponent, producing an infinite hierarchy of universality classes. This result follows because the fluctuating population produced at one stage becomes the random input to the next. The mechanism persists for both finite-variance and heavy-tailed reproduction and strongly enhances the probability of rare terminal events. Applied to multitype earthquake triggering, it provides a stationary mechanism for the anomalously large abundance of foreshock-mainshock sequences. Directed architecture can therefore control both transient amplification and critical statistics in cascade processes even when it leaves spectral stability unchanged.
\end{abstract}

\maketitle

Spectral stability determines whether a branching cascade ultimately dies out, but not how large it can become before extinction. This distinction is central to cascade dynamics: a system can be asymptotically stable and yet generate a large transient population capable of reaching rare states. Branching processes provide a natural framework for this problem, with applications ranging from population growth and epidemics to neutron multiplication, neuronal and fracture avalanches, self-exciting point processes, and earthquake triggering~\cite{harris1963theory,haccou2005branching,kimmel2015branching,pastor2015epidemic,hawkes1971spectra,ogata1988statistical}. In the scalar case, the mean fertility, or branching ratio, $R$ determines asymptotic behavior, while at criticality the total cluster-size distribution obeys $p(S)\sim 1/S^{\tau}$ with the universal mean-field exponent $\tau=3/2$ for finite offspring variance~\cite{harris1963theory}. Strong event-to-event fertility fluctuations can change this local universality class. If the expected fertility $\mu$ has a power-law tail $p(\mu)\sim 1/\mu^{1+\gamma}$ with $1<\gamma<2$, the scalar critical law becomes $p(S)\sim 1/S^{1+1/\gamma}$~\cite{saichev2005power}. In ETAS seismicity, for example, the combination of exponentially increasing productivity $\propto e^{\alpha m}$ with the Gutenberg--Richter magnitude distribution $p(m)\propto e^{-\beta m}$ produces precisely such a broad fertility distribution, with $\gamma=\beta/\alpha$~\cite{ogata1988statistical,utsu1995centenary,helmstetter2002subcritical,saichev2005power}.

Most natural cascades, however, are multichannel rather than statistically homogeneous. Events may belong to different species, epidemic stages, spatial regions, failure modes, neuronal populations, market sectors, or faulting mechanisms, with type-dependent reproduction and conversion rules. A multitype branching process is described by a non-negative matrix $\mathbf R$, where $R_{ij}$ is the mean number of type-$i$ descendants produced directly by one type-$j$ event. Multivariate Hawkes processes are the continuous-time counterpart: the integrated excitation kernels form the same matrix $\mathbf R$, and asymptotic stability is set by the spectral condition $\rho(\mathbf R)<1$~\cite{hawkes1971spectra,hawkesOakes1974cluster,bremaudMassoulie1996stability,saichev2011generating,saichev2013hierarchy,jovanovic2015cumulants,mode1971multitype,seneta2006non}. The spectral radius, however, is insensitive to how triggering pathways are organized.

Here we show that this directed architecture is an additional physical control parameter. The main surprise appears when critical components are connected successively: their number along a directed path changes the cascade-size exponent itself. For finite-variance branching, the standard $3/2$ exponent becomes $5/4$ for two successive critical components and $9/8$ for three, generating a hierarchy that approaches $1$ as the number of critical stages increases. The physical mechanism is simple: the entire fluctuating population produced by one critical stage becomes the random number of seeds injected into the next. For broad fertility distributions, the same nesting mechanism generates the corresponding $\gamma$-dependent hierarchy.

Directed architecture also controls amplification away from criticality. Feed-forward pathways can produce polynomial transient growth while leaving the eigenvalues, and hence the asymptotic stability threshold, unchanged. Their depth therefore controls how strongly activity can be amplified before it eventually decays. This separation between spectral stability, transient amplification, and critical scaling applies directly to multitype branching and Hawkes processes, and it can strongly enhance the probability of rare terminal events. In earthquake triggering, we show below that it provides a stationary mechanism for the excess abundance of foreshock--mainshock sequences relative to one-channel ETAS predictions.

The origin of these feed-forward pathways lies in the non-reciprocity of cross-channel excitation: type j may trigger type i much more efficiently than type i triggers type j. Grouping mutually interacting types into strongly connected components (SCCs) then separates the recurrent dynamics within components from the directed transfer of activity between them. The SCCs form a directed acyclic graph, yielding the standard Frobenius representation of decomposable multitype branching processes~\cite{foster1976decomposable,foster1978limit,vatutin2018decomposable}. In Frobenius form,
$\mathbf R=\mathbf A+\mathbf B$,
where $\mathbf A$ is block diagonal and contains the SCCs, while $\mathbf B$ contains the strictly feed-forward couplings between them. The spectrum of $\mathbf R$ is entirely determined by the SCC blocks in $\mathbf A$: $\mathbf A$ fixes asymptotic stability, whereas the nilpotent matrix $\mathbf B$ leaves the spectrum unchanged and transfers activity downstream.

Two distinct depths then govern the effects of the directed architecture. The \emph{feed-forward depth} $L$, defined by
$\mathbf B^{L}\neq\mathbf, ~\mathbf B^{L+1}=\mathbf0$,
is a purely structural property of the feed-forward backbone and sets the highest polynomial order of transient amplification. Critical scaling is governed instead by the \emph{critical depth} $d(j)\leq L+1$, the maximum number of critical SCCs encountered along a directed path accessible from a seed of type $j$. Thus, $L$ controls the extent of transient amplification, whereas $d(j)$ determines the critical scaling exponent. If $p(S,j)$ denotes the total cascade-size distribution from a seed of type $j$, and $\langle S\rangle_j$ its mean, then $d(j)$ fixes the tail exponent $p(S,j)\sim S^{-\tau_j}$, while the feed-forward structure controls the magnitude of the response.

\del{Near criticality, when the feed-forward links share a characteristic
strength $q$, the contribution of a directed path $P$ with $\ell(P)$
feed-forward steps and $d(P)$ critical SCCs has the general form
\begin{equation}
\langle S(j)\rangle_P
\asymp
C_P\,\frac{q^{\ell(P)}}{(1-\lambda)^{d(P)}},
\label{eq:general_tree_scaling}
\end{equation}
where $C_P$ remains nonsingular. Thus, noncritical portions of the backbone do not modify the critical exponent, but enhance the amplitude of the cascade response; the largest divergence is fixed by $d_*(j)=\max_P d(P)$.} 

This role of the critical depth $d(j)$ is quantified by the hierarchy of cascade-size exponents
\begin{equation}
\tau_j=1+
\begin{cases}
1/2^{d(j)}, & \text{finite variance},  \\
1/\gamma^{d(j)}, & 1<\gamma<2.  
\end{cases}
\label{eq:main_tau}
\end{equation}
For finite-variance branching, the standard mean-field exponent $3/2$ is recovered for $d(j)=1$, 
but becomes $5/4$ for two successive critical components and $9/8$ for three, approaching $1$ as the critical depth increases. 
For broad fertility distributions, the corresponding hierarchy is $1/\gamma^{d(j)}$.
These results follow from a direct statistical argument. Consider a finite-variance critical branching class, for which the total progeny $S$ has the asymptotic tail
\begin{equation}
\Pr(S>s)\sim s^{-1/2}.
\label{eq:scalar_survival}
\end{equation}
Now consider two consecutive critical classes. Let $S_1$ be the total progeny generated in the upstream class. Since each event in the first class has a finite probability of triggering the second class, the number of downstream seeds is, in first approximation, proportional to $S_1$. Each seed generates an independent critical progeny with the same tail as~\eqref{eq:scalar_survival}. The sum of $S_1$ such progenies has the characteristic scale
\begin{equation}
S_2\sim S_1^2,
\label{eq:sum_scale_finite}
\end{equation}
a standard consequence of summing variables with a $1/2$ survival-tail exponent~\cite{bouchaud1990anomalous,sornette2006critical}. 
From Eq.(\ref{eq:scalar_survival}),
the total cascade size $S\simeq S_2$ therefore satisfies
$\Pr(S>s)
\sim
\Pr(S_1>s^{1/2})
\sim s^{-1/4}$,
and hence $p(S)\sim S^{-5/4}$.
More generally, each additional critical class changes the characteristic scale according to $S_{\ell+1}\sim S_\ell^2$, yielding
$\Pr(S>s)\sim s^{-2^{-d}}, p(S)\sim S^{-\left(1+2^{-d}\right)}$.

\del{Consider first a finite-variance critical class.
Its total progeny $X$ has the 
complementary cumulative distribution function (CCDF) for cluster sizes
\begin{equation}
\Pr(X>x)\sim x^{-1/2}.
\label{eq:scalar_survival}
\end{equation}
Suppose that an upstream critical cascade contains $N$ events and that each event has a nonzero probability of seeding the next downstream class. The number of downstream seeds is then proportional to $N$ at leading order. Each seed initiates an independent critical progeny $X$, and the sum of $N$ such variables, whose survival tail has index $1/2$, has the characteristic scale \cite{bouchaud1990anomalous,sornette2006critical}
\begin{equation}
X_1+\cdots+X_N\sim N^2.
\label{eq:sum_scale_finite}
\end{equation}
The upstream cascade itself satisfies $\Pr(N>n)\sim n^{-1/2}$. Consequently, the combined progeny $S\sim N^2$ obeys
\begin{equation}
\Pr(S>s)\sim\Pr(N>s^{1/2})\sim s^{-1/4},
\end{equation}
and therefore $p(S)\sim S^{-5/4}$. Each additional critical channel repeats the same random-sum operation and divides the survival-tail exponent by two. After $d$ critical channels,
\begin{equation}
\Pr(S>s)\sim 1/s^{1/2^{d}},
\qquad
p(S)\sim 1/S^{1+1/2^{d}}.
\label{eq:statistical_finite}
\end{equation}}

The same argument applies to broad fertilities without modification of its logic. A scalar critical cascade now satisfies $\Pr(X>x)\sim 1/x^{1/\gamma}$. The sum of $N$ independent copies has scale $N^\gamma$ \cite{bouchaud1990anomalous,sornette2006critical}, so each additional critical channel divides the survival exponent by $\gamma$. This gives $\Pr(S>s)\sim 1/s^{1/\gamma^{d}},
p(S)\sim 1/S^{1+1/\gamma^{d}},$
which establishes the physical origin of Eqs.~(\ref{eq:main_tau}): the complete fluctuations of one branching cascade become the random number of seeds of the next.
The generating-function formulation makes the assumptions underlying
Eqs.~(\ref{eq:main_tau}) explicit
(see END MATTER).

\begin{figure}[t]
\centering
\includegraphics[width=\columnwidth]{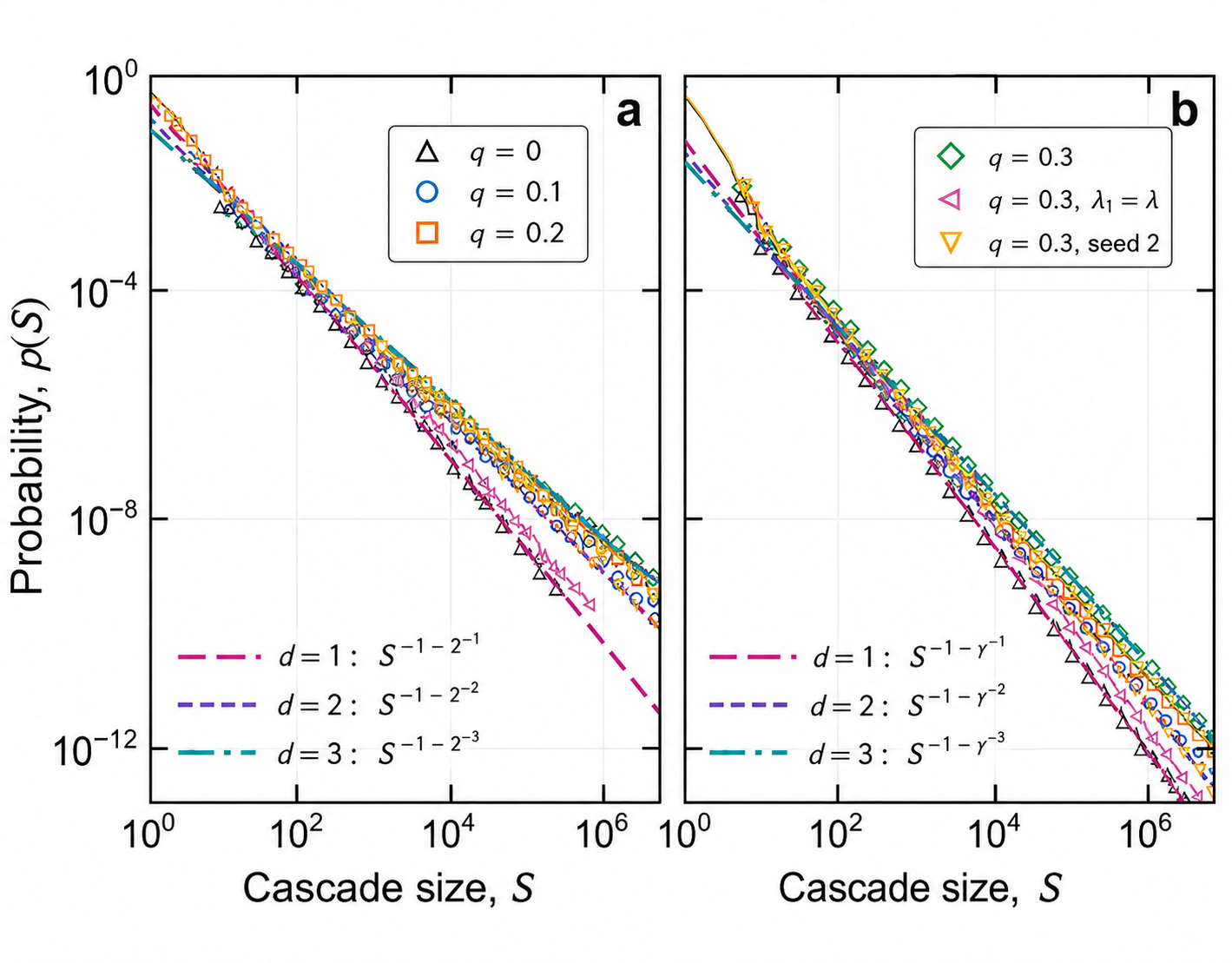}
\caption{Successive critical stages generate distinct universality classes as shown by the
total cascade-size distributions for the directed chain $3\to2\to1$ associated with matrix $\mathbf R$ (\ref{eq:compensated}). (a) Finite-variance offspring and (b) power law fertility $p(\mu)\sim 1/\mu^{1+\gamma}$. At criticality, the predictions are $p(S) \sim 1/S^{1+1/2^{d}}$ and $1/S^{1+1/\gamma^{d}}$. Away from criticality, redistributing reproduction from within-channel branching to directed cross-channel triggering leaves the ultimate scalar asymptotics unchanged but strongly enhances large cascades over a broad crossover range. Symbols are simulations; dashed lines show the theoretical critical slopes.}
\label{fig:cascade_sizes}
\end{figure}

Numerical simulations (details can be found in Supplemental Material \cite{sornette2026nilpotentsuppmat}) confirm the predicted scaling for both finite-variance offspring distributions and fertility distributions with tail index $1<\gamma<2$. We consider the representative architecture
\begin{equation}
\mathbf R=
\begin{pmatrix}
\lambda_1&q&0\\
0&\lambda_1&q\\
0&0&\lambda
\end{pmatrix},
\qquad
0\le q<\lambda<1,\quad \lambda_1<\lambda,
\label{eq:compensated}
\end{equation}
where columns label parent types and the off-diagonal elements describe the directed transitions $3\to2$ and $2\to1$, both with coupling strength $q$. As detailed in Supplemental Material \cite{sornette2026nilpotentsuppmat}, cascades initiated from different seed types probe different critical depths, $d(j)=1,2,3$. Figure~\ref{fig:cascade_sizes} shows excellent agreement between the numerical simulations and the theoretical predictions 
for all seed types and for both fertility regimes.

\del{The same directed architecture controls amplification on the subcritical side. To display it explicitly, specialize the SCCs in Eq.~(\ref{eq:AB}) to scalar classes, so that
  \begin{equation}
\mathbf A=\mathbf D
=\operatorname{diag}(\lambda_1,\ldots,\lambda_n),
\label{eq:DB}
\end{equation}
where $\lambda_i=R_{ii}$ and $B_{ij}=R_{ij}$ for $i\neq j$ is the mean
number of type-$i$ direct offspring produced by one type-$j$ event through
a cross-channel transition. The structural depth of this strictly
triangular $\mathbf B$ is the $L$ defined in Eq.~(\ref{eq:nilpotent_depth}).}

We now turn to the mean cascade size $\langle S\rangle_j$ and examine how the feed-forward structure enhances it near criticality.
For $\rho(\mathbf R)<1$, the expected total cascade size is
$\langle S(j)\rangle
=\mathbf 1^{\mathrm T}(\mathbf I-\mathbf R)^{-1}\mathbf e_j.$
The resolvent admits an exact expansion over directed paths. A path $P:i_0=j\to i_1\to\cdots\to i_r$ contributes
$W(P)= \frac{\prod_{a=0}^{r-1}B_{i_{a+1}i_a}}
{\prod_{a=0}^{r}(1-\lambda_{i_a})}$,
where $B_{i_{a+1}i_a}$ is the mean offspring number associated with the directed transition $i_a\longrightarrow i_{a+1}$.
This expression has a direct physical interpretation: the factors $B_{i_{a+1}i_a}$
quantify how efficiently activity is transmitted from one component to the next, while each factor $1/(1-\lambda_{i_a})$
measures the mean persistence within component $i_a$.
Thus, near-critical components act as long-lived amplifying stages, and the directed couplings successively transfer their enlarged populations downstream. Components that remain uniformly subcritical contribute only finite factors and therefore affect the amplitude, but not the critical singularity.

The combinatorics simplify considerably when all cross-channel couplings of the feed-forward matrix $\mathbf{B}$
in the decomposition $\mathbf R=\mathbf A+\mathbf B$ share a common strength $q>0$.
To characterize the contribution of the feed-forward structure to the mean cascade size, consider a directed path $P$ containing $\ell(P)$ feed-forward steps, and let $d(P)$ denote the number of SCCs along the path whose distance from criticality is $\epsilon\to0^+$. The contribution of such a path has the asymptotic form
$W(P)\asymp
C_P\frac{q^{\ell(P)}}{\epsilon^{d(P)}}$,
where $C_P$ remains nonsingular as $\epsilon\to0^+$. Summing over all paths accessible from a seed $j$ gives
$
\langle S(j)\rangle=\sum_P W(P)
\asymp
\frac{C_j}{\epsilon^{d(j)}}$,
where $d(j)=\max_P d(P)$.
Thus, the number of near-critical SCCs encountered along a path determines the order of the critical divergence, whereas the nonsingular prefactor $C_j$ collects the contributions from the feed-forward structure and from all paths attaining the maximum critical depth.
The structural extent of this amplification is characterized by the feed-forward depth
$L=\max_P\ell(P)$,
which sets the maximum number of feed-forward steps that can be traversed. Hence, increasing $L$ allows activity to propagate through longer feed-forward pathways and can strongly enhance the amplitude of $\langle S(j)\rangle$, without changing its critical exponent. In this sense, $L$ controls the structural capacity for transient amplification, whereas $d(j)$ determines the singularity of the mean cascade size.

\del{A homogeneous path containing $d$ simultaneously critical classes therefore gives $\langle S\rangle\asymp q^{d-1}(1-\lambda)^{-d}$.}

To elucidate the mechanism of transient amplification, we first consider the homogeneous case $\mathbf{A}=\lambda\mathbf{I}$. For a feed-forward chain of $d$ classes, the contribution reaching the terminal class after $k$ generations is
\begin{equation}
(\mathbf R^k)_{i_1i_d}
=\binom{k}{d-1}q^{d-1}\lambda^{k-d+1}
\sim\frac{q^{d-1}}{(d-1)!}k^{d-1}\lambda^{k-d+1}.
\label{eq:transient}
\end{equation}
Thus a chain with $d-1$ feed-forward steps produces a polynomial of
degree $d-1$, and the maximal degree is $L$. The eigenvalues determine
the eventual exponential decay. This is the characteristic separation
between spectral stability and transient response in non-normal
systems~\cite{neubert1997alternatives,trefethen1993hydrodynamic,trefethen1999spectra}.
In the branching context, the polynomial factors enumerate ordered
sequences of type conversions.

Figure~\ref{fig:cascade_sizes} confirms the two universality hierarchies. At strict subcriticality, the critical laws persist over an intermediate range whose extent increases on approaching the transition. For finite-variance offspring, the distribution eventually develops a cutoff. For power law fertility, it crosses over to a faster, fertility-dominated algebraic tail~\cite{saichev2005power}. Directionality nevertheless produces large amplification throughout the subcritical crossover regime.

We now apply this mechanism to earthquake triggering and precursory
sequences. The physical basis and empirical construction of a
non-normal, mechanism-resolved Hawkes--ETAS model are developed in
Ref.~\cite{sornette2026}. In this description, earthquake types
represent faulting mechanisms or other seismologically meaningful
triggering channels. Normal, strike-slip, and thrust earthquakes provide three natural
classes of triggering events, since both their magnitude statistics and
their triggering properties depend systematically on faulting style
~\cite{schorlemmer2005variations,narteau2009common,lippiello2015mechanical}.

Let $R_{ij}$ denote the baseline productivity from channel $j$ (say thrust) to
channel $i$ (say strike-slip): a type-$j$ earthquake of magnitude $m$ produces, on
average,
$
R^{\rm ETAS}_{ij}(m)
=
R_{ij}e^{\alpha(m-m_{\min})}$
direct descendants of type $i$. Assuming the GR
magnitude density
$p(m)=\beta e^{-\beta(m-m_{\min})}$, averaging this productivity over
magnitudes gives, for $\beta>\alpha$,
\begin{align}
y
&=
\left\langle e^{\alpha(m-m_{\min})}\right\rangle
=
\frac{\beta}{\beta-\alpha},
&
\overline{\mathbf R}^{\rm ETAS}
&=
y\mathbf R .
\label{eq:etas_average}
\end{align}
At the level of first moments, the existence of a broad distribution of fertilities therefore preserves the entire
multitype dynamics after the single replacement
$\mathbf R\mapsto\overline{\mathbf R}^{\rm ETAS}=y\mathbf R$: generation
profiles, resolvents, and transient polynomial amplification retain the
form of homogenous fertilities (no marks) This closure uses only the mean mark factor $y$.
Critical fluctuations are more subtle and cannot in general be obtained
by the same scalar renormalization. The complete fertility distribution
fixes the leading nonlinear singularity of the offspring generating
function. Here the productivity law and the GR distribution generate a power law fertility tail with
$\gamma=\beta/\alpha$. Consequently,
$\rho(y\mathbf R)$ determines asymptotic stability, the directed
architecture fixes the nesting depth, and $\gamma$, rather than $y$
alone, changes each nested square root into a $\gamma$th root. This
separates the structural amplification mechanism from the
fertility-dependent critical behavior.

In the conventional one-channel ETAS model \cite{ogata1988statistical,helmstetter2002subcritical}, earthquake magnitudes are independent GR draws within each triggering tree, so foreshocks arise when a small event initiates a cascade that later reaches a larger magnitude. Although this mechanism generates foreshock sequences \cite{helmstetter2003mainshocks,helmstetter2003foreshocks}, standard one-channel ETAS simulations substantially underpredict their observed abundance~\cite{mignan2014debate,petrillo2021testing,petrillo2023incorporating}. Directed multichannel branching provides a stationary mechanism for this excess: it amplifies the population of descendants, and therefore the number of opportunities for a cascade to reach the upper magnitude tail, without requiring time-dependent preparation or evolving model parameters.
Such multichannel triggering is physically natural in heterogeneous fault systems. Distinct faulting mechanisms can coexist and interact within the same tectonic environment because deformation generally cannot be accommodated by a single mode alone \cite{GabrielovKeilisBorokJackson1996,sornette2026}. A population generated in one faulting channel can therefore seed another, whose fluctuating population may in turn activate a further channel. Crucially, what is transmitted downstream is not a representative event but an entire random population. The cascade-size fluctuations produced in one channel thus become the random number of triggering opportunities supplied to the next, providing the physical origin of the amplification.

\begin{figure}[!t]
\centering
\includegraphics[width=0.95\columnwidth]{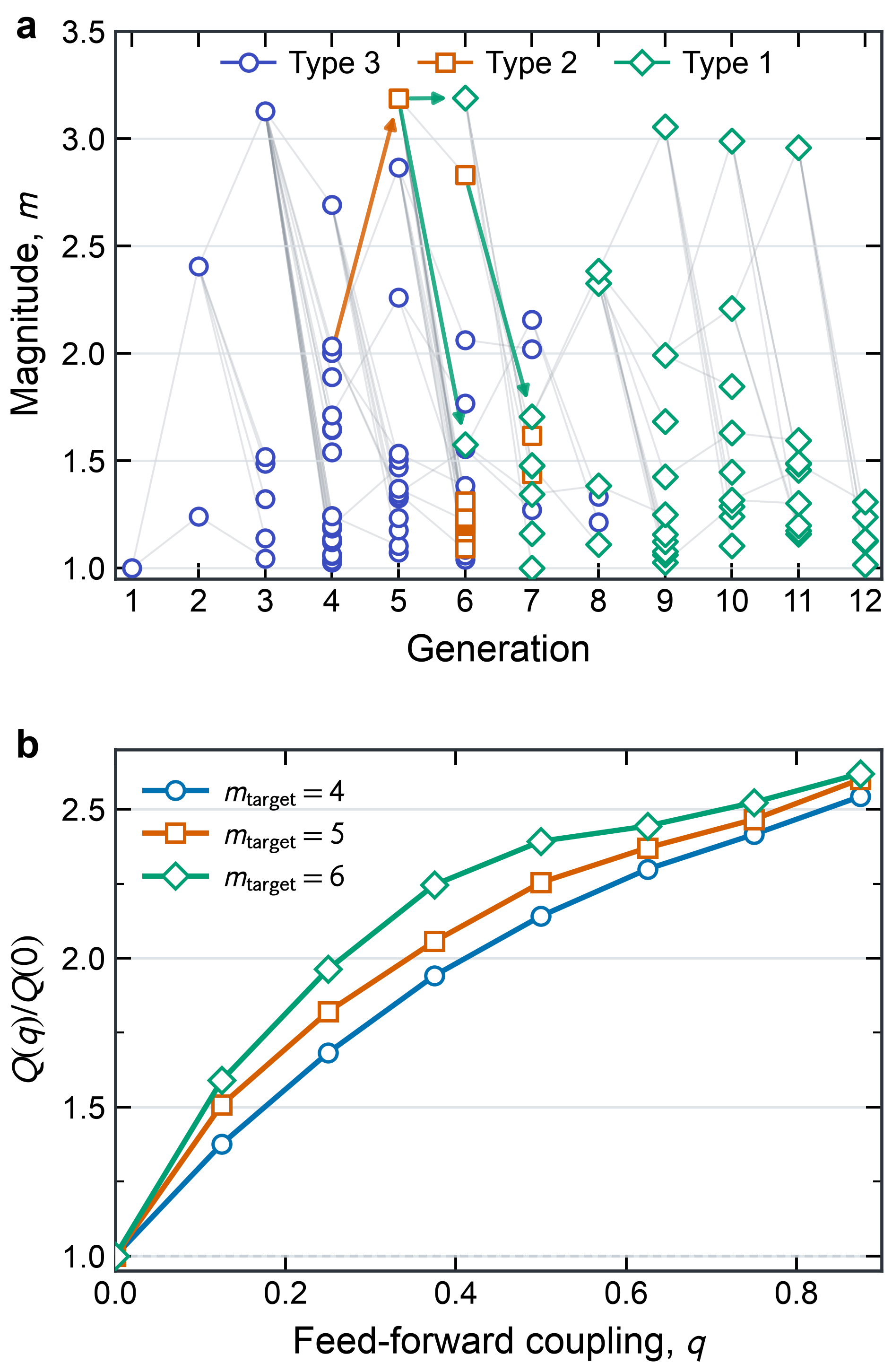}
\caption{Precursory consequences of directed coupling in a subcritical
multitype ETAS process. (a) One realization of the cascade
$3\to2\to1$. Each symbol represents an earthquake, with its generation on the
horizontal axis, its magnitude on the vertical axis, and its shape and
color identifying its triggering channel. Grey lines show the parent--offspring genealogy. Successive
activation and the overlap of different populations demonstrate how
cascade fluctuations are transmitted between channels.
(b) Enhancement factor $Q_{m_{\rm target}}(q)/Q_{m_{\rm target}}(0)$,
with $Q_{m_{\rm target}}(q)$ defined by (\ref{eq:Q}) for three target magnitudes at fixed
spectral radius $\rho(\mathbf R)=\lambda$. 
}
\label{fig:precursors}
\end{figure}

Figure~\ref{fig:precursors}(a) shows one realization of this process \add{with the matrix in Eq.(\ref{eq:compensated}) with $\lambda_1=\lambda-q$. In this case, increasing $q$ is compensated by reducing part of the diagonal
reproduction, so that the total column sum is independent of $q$.}
The cascade begins in type~3, whose
events reproduce over several generations and eventually seed type~2.
The resulting type~2 population subsequently seeds type~1, which
continues to generate events over later generations. The three
populations overlap in time: downstream activation does not require the
upstream population to have become extinct. Moreover, type~1 activity
can reappear at late generations because different branches of the
genealogy reach the downstream channel after different delays. Panel
(a) thus makes explicit the mechanism underlying the amplification:
successive random populations are transmitted through the directed
sequence $3\to2\to1$.

This transmission affects not only the mean number of descendants but
the complete cascade-size distribution. It consequently increases the
number of opportunities to sample the upper GR
magnitude tail. To quantify this effect, we initiate a cascade with a
type~3 source earthquake of magnitude $m_0=m_{\min}$ and define
\begin{equation}
Q_{m_{\rm target}}(q)
=
{\rm Pr}_q\left(m_{\max}\geq m_{\rm target}\right),
\label{eq:Q}
\end{equation}
where $m_{\max}$ is the largest magnitude reached anywhere in the
resulting multichannel cascade.

Figure~\ref{fig:precursors}(b) shows that $Q_{m_{\rm target}}(q)/Q_{m_{\rm target}}(0)$ increases strongly
with directed coupling $q$, even though the spectral radius and column count are kept
fixed. The relative enhancement is larger for the more extreme target
magnitudes. Panels~(a) and (b) therefore connect the microscopic
mechanism to its statistical consequence: successive transmission
through distinct triggering channels generates a larger and more
broadly fluctuating descendant population, which raises the probability
of reaching the remote magnitude tail. This result does not constitute
a prediction of individual earthquakes. It provides a population-level
mechanism that increases the frequency of sequences classified
retrospectively as foreshock--mainshock sequences and gives an explanation for the high foreshock rates observed in empirical
catalogs relative to standard one-channel ETAS predictions ~\cite{mignan2014debate,petrillo2021testing,petrillo2023incorporating}.

The same logic applies beyond seismicity. Whenever rare extreme events are reached through several asymmetric reproduction or conversion stages, a one-channel model underestimates the probability of large terminal events. Examples include progression between epidemic stages or host populations, mutation and phenotypic differentiation, successive failure modes in heterogeneous materials, activation across neuronal populations or cortical layers, reaction and particle-multiplication chains, and contagion across economic sectors. In each case, precursor abundance may reflect the directed organization of otherwise stationary subcritical dynamics rather than a temporal drift toward instability.


\begin{acknowledgments}
We acknowledge CINECA for computational resources through the ISCRA~C project \textit{IsCc7}. DS acknowledges support from the Center for Computational Science and Engineering at Southern University of Science and Technology.
EL acknowledges funding from the Italian Ministero dell'Universit\`a e della Ricerca under PRIN 2022 (``re-ranking of the final lists''), project 2022KWTEB7, CUP B53C24006470006. GP acknowledges support from the DyD\'eCo project and from the CPJ Chair at ENS de Lyon and CNRS.
\end{acknowledgments}

\bibliography{biblio}

\clearpage
\appendix
\section{End Matter }

Consider a directed chain of critical strongly connected components (SCC) labelled
$a=1,\ldots,d$, with triggering from class $a+1$ to class $a$.
Class $a=1$ is the terminal downstream class. Let $p_a(S)$ be the
probability that a cascade initiated by one event in class $a$ contains
$S$ events in total, including the initial event and all descendants in
class $a$ and in the downstream classes. Its probability generating
function is
\begin{equation}
H_a(z)=\sum_{S\geq1}p_a(S)z^S.
\label{eq:compact_PGF}
\end{equation}
The large-$S$ behavior of $p_a(S)$ is determined by the singularity of
$H_a(z)$ as $z\to1^-$. We therefore introduce
\begin{equation}
z=1-x,
\qquad
\epsilon_a(x)=1-H_a(1-x),
\label{eq:compact_variables}
\end{equation}
so that $x\to0^+$ probes the large-cascade limit.

We must distinguish $H_a$, which describes the complete cascade, from
the generating function of the direct offspring of one event. For a
parent in class $a+1$, let $K_{a+1}$ be the number of direct offspring
remaining in class $a+1$, and let $L_{a+1,a}$ be the number transmitted
to the next downstream class $a$. Their joint offspring generating
function is
\begin{equation}
\Phi_{a+1}(s,t)
=
\left\langle
s^{K_{a+1}}t^{L_{a+1,a}}
\right\rangle .
\label{eq:compact_offspring_PGF}
\end{equation}
The corresponding mean offspring numbers are
\begin{equation}
\left.\partial_s\Phi_{a+1}\right|_{(1,1)}
=
\langle K_{a+1}\rangle,
\qquad
q_a\equiv
\left.\partial_t\Phi_{a+1}\right|_{(1,1)}
=
\langle L_{a+1,a}\rangle .
\label{eq:compact_means}
\end{equation}
Critical reproduction within class $a+1$ means
$\langle K_{a+1}\rangle=1$, while directed transmission requires
$q_a>0$.

For finite, nonzero offspring variance, the expansion of the direct
offspring generating function around $(s,t)=(1,1)$ can be written as
\begin{align}
\Phi_{a+1}(1-u,1-v)
={}&
1-u-q_av+A_{a+1}u^2
\nonumber\\
&+C_{a+1}uv+D_{a+1}v^2
+o(u^2+uv+v^2),
\label{eq:compact_expansion}
\end{align}
where
\begin{equation}
A_{a+1}
=
\frac{1}{2}
\left.\partial_s^2\Phi_{a+1}\right|_{(1,1)}
=
\frac{\operatorname{Var}(K_{a+1})}{2}
>0.
\label{eq:compact_A}
\end{equation}
The last equality follows from
$\langle K_{a+1}\rangle=1$. The coefficients $C_{a+1}$ and
$D_{a+1}$ describe mixed and downstream offspring fluctuations,
respectively.

The exact tree equation for a cascade initiated in class $a+1$ is
\begin{equation}
H_{a+1}(z)
=
z\,
\Phi_{a+1}
\!\left(H_{a+1}(z),H_a(z)\right).
\label{eq:compact_tree}
\end{equation}
Substituting $z=1-x$, $H_{a+1}=1-\epsilon_{a+1}$, and
$H_a=1-\epsilon_a$ into Eq.~(\ref{eq:compact_tree}), and then using
the expansion (\ref{eq:compact_expansion}), gives
\begin{align}
1-\epsilon_{a+1}
={}&
(1-x)\big[
1-\epsilon_{a+1}-q_a\epsilon_a
+A_{a+1}\epsilon_{a+1}^2
\nonumber\\
&\hspace{18mm}
+C_{a+1}\epsilon_{a+1}\epsilon_a
+D_{a+1}\epsilon_a^2+\cdots
\big].
\label{eq:compact_substitution}
\end{align}
The terms $-\epsilon_{a+1}$ on the two sides cancel precisely because
the mean same-class reproduction is critical. Retaining the dominant
terms then yields
\begin{equation}
A_{a+1}\epsilon_{a+1}^2
\sim
x+q_a\epsilon_a.
\label{eq:compact_recursion}
\end{equation}
For the terminal critical class,
$A_1\epsilon_1^2\sim x$, and hence
$\epsilon_1\asymp x^{1/2}$. Since $\epsilon_1\gg x$ as $x\to0^+$,
Eq.~(\ref{eq:compact_recursion}) implies recursively
\begin{equation}
\epsilon_{a+1}
\sim
\left(\frac{q_a}{A_{a+1}}\right)^{1/2}
\epsilon_a^{1/2}.
\label{eq:compact_square_root}
\end{equation}
Thus every additional critical SCC takes one further square root of
the downstream singularity, giving
\begin{equation}
1-H_d(z)
\asymp
(1-z)^{2^{-d}}.
\label{eq:compact_nested}
\end{equation}
Standard singularity analysis~\cite{FlajoletSedgewick09} then gives
$p_d(S)\sim S^{-1-2^{-d}}$.

If the fertility distribution instead has a power-law tail with
$1<\gamma<2$, its variance diverges and the leading nonlinear term in
Eq.~(\ref{eq:compact_expansion}) is
$A_{a+1}u^\gamma$ rather than $A_{a+1}u^2$. The same substitution gives
\begin{equation}
A_{a+1}\epsilon_{a+1}^{\gamma}
\sim
x+q_a\epsilon_a,
\end{equation}
so that each additional critical SCC takes a $\gamma$th root:
$\epsilon_d\asymp x^{\gamma^{-d}}$. Consequently,
$p_d(S)\sim S^{-1-\gamma^{-d}}$. For a general seed of type $j$, $d$
is replaced by the largest accessible critical depth $d(j)$.
The coupling coefficients determine the prefactors, whereas the
critical depth and the local fertility distribution determine the
tail exponent. The complete calculation is presented in the
Supplemental Material \cite{sornette2026nilpotentsuppmat}.


\clearpage

\let\SIoldaddcontentsline\addcontentsline
\renewcommand{\addcontentsline}[3]{%
  \SIoldaddcontentsline{sitoc}{#2}{#3}%
}

\setcounter{section}{0}
\setcounter{subsection}{0}
\setcounter{equation}{0}
\setcounter{figure}{0}
\setcounter{table}{0}

\renewcommand{\theequation}{S\arabic{equation}}
\renewcommand{\thefigure}{S\arabic{figure}}
\renewcommand{\thetable}{S\arabic{table}}

\setcounter{secnumdepth}{2}
\setcounter{tocdepth}{2}

\begin{center}
{\Large\bfseries SUPPLEMENTARY INFORMATION\par}
\vspace{0.5em}
{\Large\bfseries
Directed Cascades Generate New Critical Universality Classes\par}

\vspace{1.2em}

Didier Sornette$^{1}$, Eugenio Lippiello$^{2}$, Giuseppe Petrillo$^{3}$

\vspace{0.5em}

{\small
$^{1}$Institute of Risk Analysis, Prediction and Management (Risks-X),
Academy for Advanced Interdisciplinary Studies,
Southern University of Science and Technology, Shenzhen, China\\
$^{2}$Department of Mathematics and Physics,
University of Campania ``Luigi Vanvitelli'', Italy\\
$^{3}$CNRS, ENS de Lyon, LPENSL, UMR5672,
Lyon cedex 07, Lyon, 69342, France
}
\end{center}

\vspace{1em}

\SItableofcontents

\clearpage

\section{Expected total progeny}

Consider a multitype branching process with $n$ types.
Let
$
R_{ij}
$
be the mean number of offspring of type $i$ generated by one individual of type $j$.
The process is subcritical whenever
$
\rho(\mathbf{R})<1,
$
where $\rho(\mathbf{R})$ denotes the spectral radius. If the cascade starts from one individual of type $j$, the expected total progeny (including the ancestor) is exactly
\begin{equation}
S(j)
=
{\bf 1}^{\mathrm T}
(\mathbf{I}-\mathbf{R})^{-1}
\mathbf e_j ,
\label{eq:SI_mean}
\end{equation}
where $\mathbf e_j$ is the $j$th canonical basis vector and ${\bf 1}$ denotes the vector whose entries are all equal to one.
Equation (\ref{eq:SI_mean}) is exact for every finite-type
Galton-Watson branching process with finite first moments.

\subsection{Frobenius decomposition and finite resolvent expansion}

As in the Letter, we first group mutually reachable types into strongly
connected components (SCCs) and order the SCCs topologically. The
offspring matrix then has the Frobenius decomposition
\begin{equation}
\mathbf{R}=\mathbf{A}+\mathbf{B},
\label{eq:SI_AB}
\end{equation}
where $\mathbf A$ is block diagonal and contains the internal dynamics
of the SCCs, whereas $\mathbf B$ contains only feed-forward couplings
between distinct SCCs. It is strictly block triangular and therefore
nilpotent. We use exactly the convention of the Letter: the feed-forward depth
 $L$ is defined by
\begin{equation}
\mathbf{B}^{L}\neq\mathbf0,
\qquad
\mathbf{B}^{L+1}=\mathbf0 .
\label{eq:SI_nilpotent_depth}
\end{equation}
Starting from
$
\mathbf{I}-\mathbf{R}=\mathbf{I}-\mathbf{A}-\mathbf{B},
$
we factorize
\begin{equation}
\mathbf{I}-\mathbf{R}=(\mathbf{I}-\mathbf{A})\left[\mathbf{I}-(\mathbf{I}-\mathbf{A})^{-1}\mathbf{B}
\right].
\end{equation}
Indeed,
\[
(\mathbf{I}-\mathbf{A})
\left[
\mathbf{I}-(\mathbf{I}-\mathbf{A})^{-1}\mathbf{B}
\right]
=
(\mathbf{I}-\mathbf{A})-\mathbf{B}
=
\mathbf{I}-\mathbf{R} .
\]
Taking the inverse gives
\begin{equation}
(\mathbf{I}-\mathbf{R})^{-1}
=
\left[
\mathbf{I}-(\mathbf{I}-\mathbf{A})^{-1}\mathbf{B}
\right]^{-1}
(\mathbf{I}-\mathbf{A})^{-1}.
\end{equation}
Using the Neumann identity
\[
(\mathbf{I}-\mathbf{X})^{-1}
=
\sum_{r=0}^{\infty}
X^r,
\]
with
$
\mathbf{X}=(\mathbf{I}-\mathbf{A})^{-1}\mathbf{B},
$
we obtain
\begin{equation}
(\mathbf{I}-\mathbf{R})^{-1}
=
\sum_{r=0}^{L}
\left[(\mathbf{I}-\mathbf{A})^{-1}\mathbf{B}\right]^r
(\mathbf{I}-\mathbf{A})^{-1}.
\label{eq:SI_resolvent}
\end{equation}
Multiplication by the block-diagonal matrix
$(\mathbf I-\mathbf A)^{-1}$ preserves the strictly block-triangular
structure. Hence the expansion terminates exactly at $r=L$. The SCC
blocks in $\mathbf A$ determine the spectrum and asymptotic stability,
whereas successive factors of $\mathbf B$ enumerate feed-forward
transfers between SCCs.

\subsection{Expansion over directed paths}
For explicit scalar path weights, specialize each SCC to a single type,
so that
\begin{equation}
\mathbf A=\mathbf D=\operatorname{diag}(\lambda_1,\ldots,\lambda_n).
\label{eq:SI_scalar_specialization}
\end{equation}
Equation (\ref{eq:SI_resolvent})
admits a simple genealogical interpretation.
The diagonal resolvent satisfies
\begin{equation}
(\mathbf{I}-\mathbf{D})^{-1}_{ii}
=
\frac{1}{1-\lambda_i}
\end{equation}
which represents the amplification produced
by repeated reproduction inside type $i$. Each factor of $\mathbf{B}$
corresponds instead to one directed conversion
between different types.

Consider an arbitrary directed path
\[
P:
i_0
\rightarrow
i_1
\rightarrow
\cdots
\rightarrow
i_r .
\]
Its contribution to the expected cascade size is
\begin{equation}
W(P)
=
\frac{
B_{i_1i_0}
B_{i_2i_1}
\cdots
B_{i_ri_{r-1}}
}
{
(1-\lambda_{i_0})
(1-\lambda_{i_1})
\cdots
(1-\lambda_{i_r})
}.
\label{eq:SI_path}
\end{equation}
The total expected progeny is therefore obtained
by summing over every directed path
accessible from the initial type,
\begin{equation}
S(j)
=
\sum_{P:j\rightsquigarrow i}
W(P).
\label{eq:SI_sum}
\end{equation}
Equation (\ref{eq:SI_sum})
is exact and provides a complete path representation
of the resolvent.

\subsection{Homogeneous conversion chains}
For homogeneous conversion weights, we write
\begin{equation}
\mathbf B=q\mathbf G,
\end{equation}
where \(\mathbf G\) is the binary adjacency matrix of the directed
feed-forward graph.
Suppose now that every allowed conversion has the same weight,
\[
B_{ij}=q .
\]
For a path containing $\ell$ directed conversions,
\[
\ell=r(P),
\]
Eq.~(\ref{eq:SI_path}) becomes
\begin{equation}
W(P)
=
\frac{
q^\ell
}{
\prod_{a=0}^{\ell}
(1-\lambda_{i_a})
}.
\label{eq:SI_q}
\end{equation}
The numerator depends only on the number of directed conversions,
whereas the denominator depends only on the distance from criticality
of the classes visited by the path.

\subsection{Homogeneous chain of critical SCCs}
The homogeneous multicritical limit used in the Letter corresponds to
\[
\lambda_1=\lambda_2=\cdots=\lambda.
\]
Every class visited by the longest directed path
approaches criticality simultaneously.
If this path contains $d$ critical SCCs, the exact path weight gives
\begin{equation}
S(j)
\sim
C_j
\frac{
q^{d-1}
}
{
(1-\lambda)^d
},
\label{eq:SI_eq8}
\end{equation}
where
\[
C_j
=
{\bf1}^{\mathrm T}
\mathbf G^{d-1}
\mathbf e_j
\]
counts the total weight of maximal
critical paths accessible from the seed.
This is the homogeneous specialization of the path contribution stated
in the Letter. Here $d$ is the number of critical SCCs on this path,
while the number of feed-forward links is $d-1$.

\subsection{Range of validity}

Equation (\ref{eq:SI_eq8})
does not hold for an arbitrary non-normal branching matrix.
Rather, it is the homogeneous specialization
of the exact path expansion (\ref{eq:SI_sum}).
The correct general statement is therefore:

\medskip
\emph{
For a decomposable multitype branching process,
the expected total progeny admits an exact
expansion over directed paths.
Each path contributes the product
of its conversion weights divided by the product
of the distances from criticality
of the classes visited.
Consequently,
if the initial type has access to a chain
of $d$ successive classes that become critical
simultaneously,
the expected cascade size diverges as
$(1-\lambda)^{-d}$.
For homogeneous conversion chains
this reduces to
$S\sim q^{d-1}(1-\lambda)^{-d}$.
}

\subsection{Noncritical segments and critical depth}
Assume that a path $P$ visits
$
d(P)
$
classes approaching criticality,
$
1-\lambda_i=\epsilon,
$
while the remaining
$
m
$
classes remain uniformly subcritical,
$
1-\lambda_i=\epsilon+q.
$
Since the path contains
$
n=d(P)+m
$
classes,
its length is
$
\ell=n-1.
$
Its contribution is therefore
\begin{equation}
W_{\rm full}
=
\frac{
q^{\,n-1}
}
{
\epsilon^{\,d(P)}
(\epsilon+q)^m
}.
\label{eq:SI_full}
\end{equation}
Equation (\ref{eq:SI_full})
is exact.

\subsection{Critical scaling}
Suppose first that $q$
remains finite while $\epsilon\rightarrow0$.
Then
\begin{equation}
(\epsilon+q)^m
=
q^m
\left(
1+\frac{\epsilon}{q}
\right)^m
\sim
q^m,
\end{equation}
and Eq.~(\ref{eq:SI_full}) becomes
\begin{equation}
W_{\rm full}
\sim
\frac{
q^{n-1-m}
}
{
\epsilon^{d(P)}
}.
\end{equation}
Since $d(P)=n-m$,
this simplifies to
\begin{equation}
W_{\rm full}
\sim
q^{\,d(P)-1}
\epsilon^{-d(P)}.
\label{eq:SI_scaling}
\end{equation}
Hence, every non-critical class contributes only
a finite resolvent factor,
while each critical class contributes one singular
factor $(1-\lambda)^{-1}$.
Consequently,
the critical divergence depends only on the number $d(P)$ of
simultaneously critical SCCs visited by the path. 

Maximizing over all
paths accessible from a seed of type $j$ defines the critical depth
used in the Letter,
\begin{equation}
d_*(j)=\max_{P\,\mathrm{accessible\ from}\,j}d(P).
\label{eq:SI_critical_depth}
\end{equation}
This must not be confused with $L$: the feed-forward depth depth counts all
feed-forward links, whereas $d_*(j)$ counts only critical SCCs on an
accessible path. When $d(P)=1$, all noncritical segments contribute
only finite factors and the path diverges as $(1-\lambda)^{-1}$.

\section{Matrix powers and transient amplification}

For a homogeneous scalar chain, write
$\mathbf R=\lambda\mathbf I+q\mathbf G$, where
$\mathbf G^L\neq\mathbf0$ and $\mathbf G^{L+1}=\mathbf0$. Since
$\lambda\mathbf I$ and $\mathbf G$ commute,
\begin{equation}
\mathbf{R}^k
=
\sum_{r=0}^{L}
\binom{k}{r}
q^r
\lambda^{k-r}
\mathbf{G}^r.
\label{eq:SI_matrix_power}
\end{equation}
Along the longest feed-forward chain, the dominant contribution is
\begin{equation}
(\mathbf{R}^k)_{i_{\min},i_{\max}}
=
\frac{q^L}{L!}\,
k^L\lambda^{k-L}
+\mathcal{O}(k^{L-1}\lambda^{k-L}),
\label{eq:Rk_asym}
\end{equation}

For the three-type chain $3\to2\to1$ used in the Letter, $L=2$ and
the homogeneous case gives
\begin{equation}
\mathbf{R}^k=
\begin{pmatrix}
\lambda^k &
k\lambda^{k-1}q &
\dfrac{k(k-1)}{2}\lambda^{k-2}q^2\\
0 &
\lambda^k &
k\lambda^{k-1}q\\
0 &
0 &
\lambda^k
\end{pmatrix}.
\end{equation}
The off-diagonal terms contain polynomial prefactors multiplying the exponential decay. In particular,
\begin{equation}
(\mathbf{R}^k)_{23}
\propto
k\lambda^{k},
\end{equation}
whereas
\begin{equation}
(\mathbf{R}^k)_{13}
\propto
k^2\lambda^{k}.
\end{equation}
Although every matrix element eventually decays because $\lambda<1$, these polynomial corrections produce a finite-generation maximum before the asymptotic exponential decay sets in.

In particular, there exists a value $k_{\max}$ such as off-diagonal elements present a maximum value. The calculation of $k_{\max}$ and the maximum value $y_{\max}$ is immediate in the $\lambda=\lambda_1$ case.
Taking for instance $a_{13}(k) = \frac{1}{2} q^2 k(k-1) \lambda^{k-2}$, we obtain 
$$k_{\max} = \frac{\ln \lambda - 2 - \sqrt{4 + (\ln \lambda)^2}}{2 \ln \lambda}$$
and
$$y_{\max} = \frac{q^2 (1 - 2k_{\max})}{2 \ln \lambda} \lambda^{k_{\max}-2}.$$

For $\lambda\neq\lambda_1$, the analytical expressions become more cumbersome.
\begin{equation}
\mathbf{R}^k = \begin{pmatrix} 
\lambda_1^k & k\lambda_1^{k-1}q & q^2 \frac{(k-1)\lambda_1^k - k\lambda_1^{k-1}\lambda + \lambda^k}{(\lambda_1 - \lambda)^2} \\ 
0 & \lambda_1^k & q \frac{\lambda_1^k - \lambda^k}{\lambda_1 - \lambda} \\ 
0 & 0 & \lambda^k 
\end{pmatrix}
\end{equation}

The matrix exhibits the same qualitative behavior: the off-diagonal elements initially increase, reach a maximum, and finally decay exponentially. Therefore, transient amplification is a generic consequence of directed propagation pathways rather than a peculiarity of the degenerate case.

The calculation of $k_{\max}$ in this case $\lambda_1\neq \lambda$ is more complicated. Considering $a_{13}(k) = \frac{q^2}{(\lambda - \lambda_1)^2} \left[ \lambda^k - k \lambda \lambda_1^{k-1} + (k-1) \lambda_1^k \right]$, we find that $k_{\max}$ is the solution of the transcendental equation
\begin{equation}
\left(\frac{\lambda}{\lambda_1}\right)^k\ln\lambda
=
\ln\lambda_1
+
\left(\frac{\lambda}{\lambda_1}-1\right)
\left(1+k\ln\lambda_1\right).
\end{equation}

\section{The distribution of cluster size $p(S,j)$}

\subsection{Generating functions}
Let $H_j(z)$
denote the generating function of the total progeny produced by a
cascade initiated in critical class $j$,
\begin{equation}
H_j(z)
=
\sum_{S=1}^{\infty}
p(S,j)z^S ,
\end{equation}
where $p(S,j)$ is the probability that the total cascade size equals
$S$. For a directed critical chain, we label the classes by
$a=1,\ldots,d$, starting from the terminal downstream class $a=1$.
The index $j$ is instead reserved for a generic seed type. Near the singular point $z=1$, we introduce
\begin{equation}
x=1-z,
\qquad
\epsilon_a
=
1-H_a(z).
\end{equation}
The asymptotic behavior of $\epsilon_a(x)$ determines the tail of the
cascade-size distribution through standard singularity analysis.


\subsection{Joint offspring generating function and its expansion}

The total-cascade generating function must be distinguished from the
generating function of the direct offspring of one event. For an event
in class $a+1$, let $K_{a+1}$ be the number of direct offspring that
remain in class $a+1$, and let $L_{a+1,a}$ be the number transmitted to
the next downstream class $a$. Their joint offspring generating function
is
\begin{equation}
\Phi_{a+1}(s,t)
=\left\langle s^{K_{a+1}}t^{L_{a+1,a}}\right\rangle .
\label{eq:SI_joint_offspring_PGF}
\end{equation}
Its first derivatives at $(1,1)$ are the corresponding mean offspring
numbers. Critical same-class reproduction and nonzero feed-forward
coupling therefore mean
\begin{equation}
\left.\partial_s\Phi_{a+1}\right|_{(1,1)}=1,
\qquad
q_a\equiv
\left.\partial_t\Phi_{a+1}\right|_{(1,1)}>0.
\label{eq:SI_critical_means}
\end{equation}

Assuming finite second moments, we set $s=1-u$ and $t=1-v$ and expand
around $(1,1)$:
\begin{align}
\Phi_{a+1}(1-u,1-v)
={}&1-u-q_av+A_{a+1}u^2
\nonumber\\
&+C_{a+1}uv+D_{a+1}v^2
\nonumber\\
&+o(u^2+uv+v^2).
\label{eq:SI_PGF}
\end{align}
The curvature in the same-class direction is
\begin{align}
A_{a+1}
&=\frac12\left.\partial_s^2\Phi_{a+1}\right|_{(1,1)}
\nonumber\\
&=\frac12\left\langle K_{a+1}(K_{a+1}-1)\right\rangle
\nonumber\\
&=\frac{\operatorname{Var}(K_{a+1})}{2}>0,
\label{eq:SI_A_definition}
\end{align}
where the last equality uses $\langle K_{a+1}\rangle=1$. The
coefficients $C_{a+1}$ and $D_{a+1}$ encode mixed and downstream
offspring fluctuations. No Poisson assumption is required.


\subsection{Terminal critical class}

Consider first the terminal critical class.
Its generating function satisfies
\begin{equation}
H_1(z)=z\Phi_1(H_1(z)).
\label{eq:SI_terminal_tree}
\end{equation}
Substituting
$
H_1=1-\epsilon_1,~
z=1-x,
$
and using the one-variable specialization of
Eq.~(\ref{eq:SI_PGF}) gives
$
1-\epsilon_1
=
(1-x)
\left(
1-\epsilon_1
+A_1\epsilon_1^2
+o(\epsilon_1^2)
\right).
$
Expanding to leading order yields
$
x
\sim
A_1
\epsilon_1^2,
$
or equivalently
\begin{equation}
\epsilon_1
\sim
x^{1/2}.
\label{eq:SI_eps1}
\end{equation}
This is the classical square-root singularity responsible for the
well-known $S^{-3/2}$ critical branching law.


\subsection{Recursive critical classes}

Suppose now that critical class $a+1$
feeds class $a$ through a non-zero coupling.
The exact tree equation is
\begin{equation}
H_{a+1}(z)
=z\Phi_{a+1}\!\left(H_{a+1}(z),H_a(z)\right).
\label{eq:SI_upstream_tree}
\end{equation}
The first argument accounts for offspring remaining in class $a+1$;
the second accounts for offspring transmitted to class $a$, each of
which generates its own complete downstream cascade. Substitution of
$z=1-x$ and $H_b=1-\epsilon_b$ into
Eq.~(\ref{eq:SI_upstream_tree}), followed by use of
Eq.~(\ref{eq:SI_PGF}), cancels the term linear in
$\epsilon_{a+1}$ because same-class reproduction is critical. The
leading balance is
\begin{equation}
A_{a+1}\epsilon_{a+1}^2
\sim x+q_a\epsilon_a.
\label{eq:SI_recursion}
\end{equation}
The omitted terms
$\epsilon_{a+1}\epsilon_a$, $\epsilon_a^2$, and
$x\epsilon_{a+1}$ are asymptotically smaller. Since
Eq.~(\ref{eq:SI_eps1}) implies $\epsilon_1\gg x$, induction gives
$\epsilon_a\gg x$ at every subsequent level. Hence,
\begin{equation}
\epsilon_{a+1}
\sim
\left(\frac{q_a}{A_{a+1}}\right)^{1/2}
\epsilon_a^{1/2}.
\label{eq:SI_square_root_recursion}
\end{equation}
Each additional critical feed-forward class therefore takes one further
square root of the singularity generated downstream.


\subsection{Induction}

Starting from
$
\epsilon_1
\sim
x^{1/2},
$
successive application of the previous relation gives
$
\epsilon_2
\sim
x^{1/4},
$
$
\epsilon_3
\sim
x^{1/8},
$
and by induction,
\begin{equation}
\epsilon_d
\sim
x^{2^{-d}}.
\label{eq:SI_epsd}
\end{equation}
Therefore each additional critical class halves the singularity
exponent.


\subsection{Singularity analysis}

By the Transfer Theorem of Flajolet and Sedgewick
\cite[Theorem VI.1]{FlajoletSedgewick09}, if
$
1-H(z)\sim(1-z)^\alpha,
$
then, using
$
\Gamma(-\alpha)
=
-\frac{\Gamma(1-\alpha)}{\alpha},
$
the coefficients satisfy
\begin{equation}
p(S)
=
[z^S]H(z)
\sim
\frac{\alpha}{\Gamma(1-\alpha)}
S^{-1-\alpha}.
\end{equation}
In our specific case
\begin{equation}
[z^S]H_d(z)
\sim
C_d
S^{-1-2^{-d}},
\end{equation}
where $C_d$ is a non-universal constant. Consequently,
\begin{equation}
p(S,d)
\sim
C_d
S^{-\tau_d},
\end{equation}
with
\begin{equation}
\tau_d
=
1+2^{-d}.
\label{eq:SI_tau}
\end{equation}
This establishes the finite-variance hierarchy stated in the Letter.


\subsection{Conditions of validity}

The above derivation relies only on the following assumptions:
\begin{enumerate}

\item
the process is decomposable,
or equivalently the critical classes can be arranged in feed-forward
order;

\item
each critical class possesses finite,
strictly positive offspring variance;

\item
successive critical classes are connected by non-zero coupling;

\item
the entire chain is accessible from the initial type;

\item
all classes along the chain become critical simultaneously;

\item
the offspring generating functions are analytic around the critical
point.

\end{enumerate}
Remarkably,
Poisson statistics are never required.


\subsection{Several accessible paths}

Suppose now that several directed critical chains are accessible from
the initial seed. Let $d(P)$
denote the number of critical SCCs visited by path $P$; uniformly
subcritical SCCs on the same path are not counted.
The dominant asymptotic behavior is governed by the deepest accessible
critical chain,
\begin{equation}
d_*(j)
=
\max_P
d(P),
\end{equation}
provided at least one maximal-depth path has non-zero weight.
Therefore
\begin{equation}
p(S,j)
\sim
S^{-1-2^{-d_*(j)}}.
\end{equation}
All shorter chains contribute only subleading corrections.


\subsection{Universality statement}

The previous derivation establishes the following result.

\medskip

\noindent
\textbf{Universality theorem.}

\smallskip

\emph{
For a decomposable critical multitype Galton--Watson branching process
with finite, non-zero offspring variances,
the total progeny distribution generated by a seed having access to a
chain of $d$ successive critical classes obeys
\begin{equation}
p(S)
\sim
S^{-1-2^{-d}}.
\end{equation}
The exponent depends exclusively on the number of successive critical
classes along the longest accessible feed-forward chain.
Offspring distributions,
coupling strengths,
and microscopic details modify only the non-universal prefactor.
}

\medskip

Equation~(\ref{eq:SI_tau}) therefore represents a universality law for
decomposable finite-variance multitype branching processes.
It should not be interpreted as applying to arbitrary critical
multitype branching matrices.
In particular,
irreducible critical multitype branching processes generally recover the
ordinary mean-field exponent $\tau=\frac32$, whereas the hierarchy
$
\tau_d
=
1+2^{-d}
$
originates from the successive nesting of critical feed-forward classes.

Equation~(\ref{eq:SI_tau}) is therefore universal only within the class of
finite-variance \emph{decomposable} critical multitype branching processes.
The distinction is structural. In a decomposable process, the critical
classes can be ordered so that activity passes only downstream from one class
to the next. Each additional critical class in the longest accessible chain
adds another level of critical nesting, giving
\[
\tau_d=1+2^{-d}.
\]

For example, two critical classes coupled only in one direction may have a
mean offspring matrix of the form
\[
\mathbf R_{\rm dec}=
\begin{pmatrix}
1 & q\\
0 & 1
\end{pmatrix},
\qquad q>0.
\]
A seed that can access both critical classes has $d=2$ and hence
$\tau_2=5/4$. By contrast, an irreducible critical matrix such as
\[
\mathbf R_{\rm irr}=
\begin{pmatrix}
1/2 & 1/2\\
1/2 & 1/2
\end{pmatrix}
\]
has mutually communicating types and constitutes a single critical class;
it therefore exhibits the ordinary finite-variance mean-field exponent
$\tau=3/2$. Thus the exponents $\tau_d<3/2$ for $d>1$ are not a generic
property of critical multitype branching, but arise specifically from a
feed-forward chain of successive critical classes.

\subsection*{Crossover from a decomposable to an irreducible critical process}

Let us introduce a weak backward coupling $\eta>0$ while keeping the matrix critical. 
For this, we retune the diagonal reproduction rates and write
\begin{equation}
\mathbf R_\eta=
\begin{pmatrix}
1-a & q\\
\eta & 1-a
\end{pmatrix},
\qquad
a=\sqrt{q\eta}.
\label{eq:Reta}
\end{equation}
The two eigenvalues are
\begin{equation}
\lambda_+=1,
\qquad
\lambda_-=1-2\sqrt{q\eta},
\end{equation}
so that $\rho(\mathbf R_\eta)=1$ for every $\eta>0$. At the same time,
$\mathbf R_\eta$ is irreducible for every $\eta>0$, because the two classes
communicate in both directions. Thus $\eta=0$ is a singular limit: exactly at
$\eta=0$ the system contains two successive critical classes, whereas for any
$\eta>0$ it contains a single irreducible critical class.

To estimate the crossover scale, let $\varepsilon=1-z$,
and denote by $x$ and $y$ the deficits of the total-progeny generating
functions from unity. Expanding the offspring generating functions about the
critical fixed point, finite non-zero offspring variances give, to leading
order,
\begin{align}
a x-qy
&\simeq
\varepsilon-c_1x^2,
\label{eq:cross1}\\
a y-\eta x
&\simeq
\varepsilon-c_2y^2,
\label{eq:cross2}
\end{align}
where $c_1,c_2>0$ are non-universal constants determined by the second moments
of the offspring distributions. Only the powers of $\varepsilon$ are needed
for the following scaling argument.

In the strictly decomposable limit $\eta=0$, hence $a=0$,
Eq.~\eqref{eq:cross2} gives
\begin{equation}
y\sim \varepsilon^{1/2}.
\end{equation}
Substituting this result into Eq.~\eqref{eq:cross1} yields
\begin{equation}
x^2\sim qy
\sim q\varepsilon^{1/2},
\end{equation}
and therefore
\begin{equation}
x\sim \varepsilon^{1/4}.
\end{equation}
The singularity $1-H(z)\sim(1-z)^{1/4}$ implies
\begin{equation}
p(S)\sim S^{-5/4},
\end{equation}
as expected for a chain of two critical classes.

For $0<\eta\ll1$, the decomposable scaling remains valid as long as the terms
introduced by the critical retuning are asymptotically negligible. In
Eq.~\eqref{eq:cross2}, the first such correction is the linear term $a y$.
Using the decomposable estimate $y\sim\varepsilon^{1/2}$, this term becomes
comparable to the leading term $\varepsilon$ when
\begin{equation}
a\varepsilon^{1/2}\sim\varepsilon.
\end{equation}
Thus, the reducible fixed-point scaling is observed for
$\varepsilon\gg a^2=q\eta$. In contrast, sufficiently close to $z=1$, the
weak backward coupling is resolved and the process crosses over to the
irreducible critical behavior.

Coefficient asymptotics probe the singularity on the scale
$1-z\sim S^{-1}$. The cross-over occurs for  $\varepsilon_\times = a^2=q\eta$,
which corresponds to a
cluster-size crossover
\begin{equation}
S_\times\sim\frac{1}{q\eta}
\end{equation}
up to a non-universal multiplicative constant. For fixed $q>0$, the extent of
the intermediate decomposable regime consequently diverges as
\begin{equation}
S_\times\propto\eta^{-1}
\qquad (\eta\to0^+).
\end{equation}

The resulting crossover for the distribution of cluster sizes  has the scaling form
\begin{equation}
p(S)\sim
\begin{cases}
S^{-5/4}, & 1\ll S\ll S_\times,\\[1mm]
S^{-3/2}, & S\gg S_\times,
\end{cases}
\qquad
S_\times\propto\eta^{-1}.
\end{equation}
Thus an arbitrarily weak backward coupling does not immediately erase the
$5/4$ law. Instead, after retuning the process to criticality, the
feed-forward exponent survives as an intermediate asymptotic over a range of
cluster sizes that grows as an inverse power of the feedback strength. The
true $S\to\infty$ asymptotic exponent is nevertheless $3/2$ for every fixed
$\eta>0$, because the two classes then form a single irreducible critical
class.

\section{The total progeny in the ETAS model}
\label{sec:SI_ETAS}

In this section we derive the marked counterpart of the mean-resolvent
formula used in the Letter.
The derivation closely parallels that of the unmarked branching process,
the only additional ingredient being the magnitude-dependent fertility of
each event.

\subsection{Marked multitype ETAS process}

Consider a multitype ETAS model with $n$ earthquake types.
The mean number of direct aftershocks generated by an earthquake of
magnitude $m$ and type $j$ is assumed to factorize into a
magnitude-dependent productivity and a type-transition matrix,
\begin{equation}
\mathbf{a}(m)
=
a(m)\mathbf{R}\mathbf e_j,
\label{eq:SI_marked_vector}
\end{equation}
where
\begin{equation}
a(m)
=
e^{\alpha(m-m_{\min})},
\label{eq:SI_productivity}
\end{equation}
is the standard ETAS productivity law,
$\mathbf{R}$ is the multitype triggering matrix,
and $\mathbf e_j$ denotes the unit vector corresponding to the initiating type.

Throughout this section, we assume that descendant magnitudes are
independently drawn from the Gutenberg--Richter distribution,
\begin{equation}
p(m)
=
\beta
e^{-\beta(m-m_{\min})},
\qquad
\beta>\alpha,
\label{eq:SI_GR}
\end{equation}
so that the average productivity is finite.

\subsection{Average productivity}

The mean productivity of an earthquake is
\begin{equation}
y
=
\langle a(m)\rangle
=
\int_{m_{\min}}^\infty
a(m)p(m)\,dm .
\end{equation}
Substituting Eqs.~(\ref{eq:SI_productivity})
and (\ref{eq:SI_GR}) gives
\begin{align}
y
&=
\beta
\int_{m_{\min}}^\infty
e^{\alpha(m-m_{\min})}
e^{-\beta(m-m_{\min})}
dm
\\
&=
\beta
\int_0^\infty
e^{-(\beta-\alpha)x}
dx
=
\frac{\beta}{\beta-\alpha}.
\end{align}
Consequently,
the average offspring matrix for every generation after the initiating
earthquake is simply
\begin{equation}
\mathbf R_{\rm ETAS}
=
y\mathbf R.
\label{eq:SI_mean_matrix}
\end{equation}
At the level of first moments, the marked ETAS process therefore differs
from the corresponding unmarked branching process only through the
multiplicative factor $y$. This reduction does not determine higher
moments or the singular part of the total-progeny distribution, which
retain information about the full mark-dependent fertility law.

\subsection{Expected cascade size}

Consider an initiating earthquake of fixed magnitude $m_0$
and type $j$.
Generation zero consists only of the initiating event,
$N^{(0)}=1$.
The expected number of first-generation aftershocks is
\begin{equation}
\langle N^{(1)}\rangle
=
a(m_0)
{\bf1}^{\mathrm T}
\mathbf{R}\mathbf e_j .
\end{equation}
After the first generation,
magnitudes are independently sampled from the
Gutenberg--Richter distribution.
Consequently,
each subsequent generation acquires an average productivity factor $y$.

The expected number of second-generation events is therefore
\begin{equation}
\langle N^{(2)}\rangle
=
a(m_0)
y
{\bf1}^{\mathrm T}
\mathbf{R}^2\mathbf e_j .
\end{equation}
Proceeding recursively,
\begin{equation}
\langle N^{(k)}\rangle
=
a(m_0)
y^{k-1}
{\bf1}^{\mathrm T}
\mathbf{R}^k\mathbf e_j,
\qquad
k\ge1.
\label{eq:SI_generation}
\end{equation}
Equation (\ref{eq:SI_generation}) is the marked analogue of the
generation expansion of an ordinary multitype branching process.

\subsection{Exact resolvent formula}

The expected total number of earthquakes generated by the initiating
event, including the root earthquake itself, is
\begin{equation}
S_{\rm ETAS}(j,m_0)
=
1+
\sum_{k=1}^{\infty}
\langle N^{(k)}\rangle .
\end{equation}
Substituting Eq.~(\ref{eq:SI_generation}) gives
\begin{equation}
S_{\rm ETAS}(j,m_0)
=
1+
a(m_0)
\sum_{k=1}^{\infty}
y^{k-1}
{\bf1}^{\mathrm T}
\mathbf{R}^k\mathbf e_j .
\end{equation}
Extracting one factor of $\mathbf{R}$,
\begin{equation}
S_{\rm ETAS}(j,m_0)
=
1+
a(m_0)
{\bf1}^{\mathrm T}
\mathbf{R}
\left(
\sum_{k=0}^{\infty}
(y\mathbf{R})^k
\right)
\mathbf e_j .
\end{equation}
Whenever $\rho(y\mathbf{R})<1$, the geometric series converges,
\begin{equation}
\sum_{k=0}^{\infty}
(y\mathbf{R})^k
=
(\mathbf{I}-y\mathbf{R})^{-1},
\end{equation}
leading to the exact expression
\begin{equation}
S_{\rm ETAS}(j,m_0)
=
1+
a(m_0)
{\bf1}^{\mathrm T}
\mathbf{R}
(\mathbf{I}-y\mathbf{R})^{-1}
\mathbf e_j.
\label{eq:SI_ETAS_resolvent_first}
\end{equation}
Equation (\ref{eq:SI_ETAS_resolvent_first})
is the exact marked counterpart of the resolvent formula
\begin{equation}
S(j)
=
{\bf1}^{\mathrm T}
(\mathbf{I}-\mathbf{R})^{-1}
\mathbf e_j
\end{equation}
derived for the unmarked process.

\subsection{Homogeneous nilpotent chain}

Consider now the homogeneous decomposition
\begin{equation}
\mathbf{R}
=
\lambda \mathbf{I}
+
\mathbf{B},
\end{equation}
where, consistently with Eq.~(\ref{eq:SI_nilpotent_depth}),
$
\mathbf{B}^{L}\neq\mathbf0,~
\mathbf{B}^{L+1}=\mathbf0.
$
We write the nilpotent feed-forward matrix as
\begin{equation}
\mathbf B=q\mathbf G,
\end{equation}
where $\mathbf G$ is a strictly triangular adjacency matrix satisfying
$
\mathbf G^L\neq\mathbf0,~
\mathbf G^{L+1}=\mathbf0.
$
Therefore,
\begin{equation}
\mathbf I-y\mathbf R
=
(1-y\lambda)\mathbf I-yq\mathbf G.
\end{equation}
Factoring out the diagonal term,
\begin{equation}
(\mathbf{I}-y\mathbf{R})^{-1}
=
\frac{1}{1-y\lambda}
\left[
\mathbf{I}-
\frac{yq}{1-y\lambda} \mathbf G
\right]^{-1}.
\end{equation}
Using again the Neumann expansion,
\[
(\mathbf{I}-X)^{-1}
=
\sum_{r=0}^{\infty}
X^r,
\]
and exploiting the nilpotency of $\mathbf G$,
\begin{equation}
(\mathbf{I}-y\mathbf{R})^{-1}
=
\sum_{r=0}^{L}
\frac{(yq)^r}
{(1-y\lambda)^{r+1}}
\mathbf G^r.
\label{eq:SI_ETAS_neumann}
\end{equation}
Substituting into Eq.~(\ref{eq:SI_ETAS_resolvent_first}) yields
\begin{equation}
S_{\rm ETAS}(j,m_0)
=
1+
a(m_0)
\sum_{r=0}^{L}
\frac{(yq)^r}
{(1-y\lambda)^{r+1}}
{\bf1}^{\mathrm T}
\mathbf{R}\mathbf G^r\mathbf e_j.
\label{eq:SI_general_ETAS}
\end{equation}

\subsection{Three-type chain}

For the elementary chain
$
3\rightarrow2\rightarrow1,
$
the structural depth is $L=2$.
Since  $\mathbf G^3=0$, Eq.~(\ref{eq:SI_ETAS_neumann})
reduces to
\begin{equation}
(\mathbf{I}-y\mathbf{R})^{-1}
=
\frac{\mathbf{I}}{1-y\lambda}
+
\frac{yq}{(1-y\lambda)^2}\mathbf G
+
\frac{y^2q^2}{(1-y\lambda)^3}\mathbf G^2.
\end{equation}
If the cascade starts from the upstream type ($j=3$) and its initiating
magnitude is also averaged over the Gutenberg--Richter law, one obtains
\begin{equation}
\left\langle S_{\rm ETAS}(3)\right\rangle
=
\left[
\frac{1}{1-y\lambda}
+
\frac{yq}{(1-y\lambda)^2}
+
\frac{y^2q^2}{(1-y\lambda)^3}
\right].
\label{eq:SI_three_chain}
\end{equation}
This expression is exact. For a fixed initiating magnitude $m_0$, the
corresponding formula contains the root-dependent prefactor derived
below in Eq.~(\ref{eq:SI_ETAS_resolvent}).

\subsection{Equivalent resolvent representation}

The expected total number of earthquakes generated by an initiating
event of type $j$ and magnitude $m_0$, including the initiating event
itself, is
\begin{equation}
S_{\rm ETAS}(j,m_0)
=
1+
\sum_{k=1}^{\infty}
\langle N^{(k)}\rangle .
\end{equation}
Using Eq.~(\ref{eq:SI_generation}) we obtain
\begin{equation}
S_{\rm ETAS}(j,m_0)
=
1+
a(m_0)
\sum_{k=1}^{\infty}
y^{k-1}
{\bf1}^{\mathrm T}
\mathbf{R}^k\mathbf e_j .
\end{equation}
Extracting one factor of $\mathbf{R}$ gives
\begin{equation}
S_{\rm ETAS}(j,m_0)
=
1+
a(m_0)
{\bf1}^{\mathrm T}
\mathbf{R}
\left(
\sum_{k=0}^{\infty}
(y\mathbf{R})^k
\right)
\mathbf e_j .
\end{equation}
Since $\rho(y\mathbf{R})<1$,
\begin{equation}
\sum_{k=0}^{\infty}
(y\mathbf{R})^k
=
(\mathbf{I}-y\mathbf{R})^{-1},
\end{equation}
and therefore
\begin{equation}
S_{\rm ETAS}(j,m_0)
=
1+
a(m_0)
{\bf1}^{\mathrm T}
\mathbf{R}(\mathbf{I}-y\mathbf{R})^{-1}\mathbf e_j .
\label{eq:SI_ETAS_first}
\end{equation}
Introducing the magnitude-averaged ETAS offspring matrix
\begin{equation}
\mathbf R_{\rm ETAS}
=
y\mathbf R,
\end{equation}
we use the matrix identity
\begin{equation}
(\mathbf I-\mathbf R_{\rm ETAS})^{-1}
-
\mathbf I
=
\mathbf R_{\rm ETAS}
(\mathbf I-\mathbf R_{\rm ETAS})^{-1}.
\end{equation}
We immediately obtain
\begin{equation}
\mathbf R
(\mathbf I-y\mathbf R)^{-1}
=
\frac{1}{y}
\left[
(\mathbf I-y\mathbf R)^{-1}
-
\mathbf I
\right].
\end{equation}
Hence, the exact resolvent formula can be written as
\begin{equation}
S_{\rm ETAS}(j,m_0)
=
1
-
\frac{a(m_0)}{y}
+
\frac{a(m_0)}{y}
{\bf1}^{\mathrm T}
(\mathbf{I}-y\mathbf{R})^{-1}
\mathbf e_j.
\label{eq:SI_ETAS_resolvent}
\end{equation}
Equation~(\ref{eq:SI_ETAS_resolvent}) is the exact marked analogue of
\[
S(j)
=
{\bf1}^{\mathrm T}
(\mathbf{I}-\mathbf{R})^{-1}
\mathbf e_j
\]
for ordinary multitype branching processes. The only modification is the
replacement of the branching matrix $\mathbf{R}$ by its effective marked
counterpart $y\mathbf{R}$, together with the prefactor associated with the
initiating earthquake.

\subsection{Critical scaling}

We now investigate the behavior of Eq.~(\ref{eq:SI_ETAS_resolvent})
close to the ETAS critical point,
\begin{equation}
y\lambda\rightarrow1^-.
\end{equation}
For the homogeneous decomposition
\begin{equation}
\mathbf{R}
=
\lambda\mathbf I+q\mathbf G,
\qquad
\mathbf G^L\neq\mathbf0,
\quad
\mathbf G^{L+1}=\mathbf0,
\end{equation}
Eq.~(\ref{eq:SI_ETAS_neumann}) gives
\begin{equation}
(\mathbf{I}-y\mathbf{R})^{-1}
=
\sum_{r=0}^{L}
\frac{(yq)^r}
{(1-y\lambda)^{r+1}}
\mathbf G^r.
\end{equation}
Substituting this expression into
Eq.~(\ref{eq:SI_ETAS_resolvent}) yields
\begin{equation}
S_{\rm ETAS}(j,m_0)
=
1-\frac{a(m_0)}{y}
+
\frac{a(m_0)}{y}
\sum_{r=0}^{L}
\frac{(yq)^r}
{(1-y\lambda)^{r+1}}
{\bf1}^{\mathrm T}
\mathbf G^r\mathbf e_j.
\label{eq:SI_scaling1}
\end{equation}
In this homogeneous critical limit, let $d_*(j)$ be the number of
critical SCCs on the deepest path accessible from the initiating type
$j$. Such a path contains $d_*(j)-1$ feed-forward links, and
\[
{\bf1}^{\mathrm T}
\mathbf G^{d_*(j)-1}
\mathbf e_j
>0,
\]
The leading singular contribution accessible from $j$ is therefore
\begin{equation}
S_{\rm ETAS}(j,m_0)
\sim
\frac{a(m_0)}{y}
\frac{(yq)^{d_*(j)-1}}
{(1-y\lambda)^{d_*(j)}}
{\bf1}^{\mathrm T}
\mathbf G^{d_*(j)-1}
\mathbf e_j.
\end{equation}
Introducing the geometric factor
\begin{equation}
C_j
=
{\bf1}^{\mathrm T}
\mathbf G^{d_*(j)-1}
\mathbf e_j,
\end{equation}
we obtain
\begin{equation}
S_{\rm ETAS}(j,m_0)
\sim
C_j\,
a(m_0)\,
y^{d_*(j)-2}
q^{d_*(j)-1}
(1-y\lambda)^{-d_*(j)}.
\label{eq:SI_ETAS_scaling}
\end{equation}
Equation~(\ref{eq:SI_ETAS_scaling}) is the fixed-root, marked analogue
of the homogeneous path scaling in the Letter.
The only modification with respect to the unmarked process is the
replacement
\[
1-\lambda
\longrightarrow
1-y\lambda,
\]
together with the multiplicative factor
$a(m_0)y^{d_*(j)-2}$, which depends on the productivity of the initiating
earthquake but leaves the critical exponent unchanged.

\subsection{Averaging over the initiating magnitude}

The previous expression refers to an initiating earthquake with fixed
magnitude $m_0$.
If the initiating magnitude itself is assumed to be drawn from the
Gutenberg--Richter distribution, its average productivity is
\begin{equation}
\langle a(m_0)\rangle =
\int_{m_{\min}}^\infty
a(m_0)p(m_0)\,dm_0
=
\frac{\beta}{\beta-\alpha}
=
y.
\end{equation}
Therefore
\begin{equation}
\langle a(m_0)\rangle=y.
\end{equation}
Taking the average of
Eq.~(\ref{eq:SI_ETAS_resolvent})
over the initiating magnitude immediately gives
\begin{align}
\left\langle
S_{\rm ETAS}(j,m_0)
\right\rangle
&=
1-\frac{\langle a(m_0)\rangle}{y}
+
\frac{\langle a(m_0)\rangle}{y}
{\bf1}^{\mathrm T}
(\mathbf{I}-y\mathbf{R})^{-1}
\mathbf e_j
\\
&=
{\bf1}^{\mathrm T}
(\mathbf{I}-y\mathbf{R})^{-1}
\mathbf e_j.
\end{align}
Hence
\begin{equation}
\left\langle
S_{\rm ETAS}(j)
\right\rangle
=
{\bf1}^{\mathrm T}
(\mathbf{I}-y\mathbf{R})^{-1}
\mathbf e_j.
\label{eq:SI_ETAS_average}
\end{equation}
Remarkably, after averaging over the initiating magnitude, the ETAS
expression becomes formally identical to the resolvent formula of the
unmarked multitype branching process, the only difference being the
replacement of the branching matrix $\mathbf{R}$ by its effective counterpart
$y\mathbf{R}$.

\subsection{General ETAS scaling theorem}

The previous derivation establishes the following result.

\medskip

\noindent
\textbf{ETAS mean-scaling theorem.}

\smallskip

\emph{
Consider a decomposable multitype ETAS process with Gutenberg--Richter
distributed magnitudes and productivity law
$a(m)=e^{\alpha(m-m_{\min})}$.
If an initiating event of type $j$ has critical depth $d_*(j)$, the
expected total number of events
generated by an earthquake of magnitude $m_0$ obeys
\[
S_{\rm ETAS}(j,m_0)
\sim
C_j\,
a(m_0)\,
y^{d_*(j)-2}
q^{d_*(j)-1}
(1-y\lambda)^{-d_*(j)},
\]
where
\[
y=\frac{\beta}{\beta-\alpha},
\]
and $C_j$ depends only on the topology of the accessible directed
chains.
After averaging over the initiating magnitude, it holds that
\[
\left\langle
S_{\rm ETAS}(j)
\right\rangle
=
{\bf1}^{\mathrm T}
(\mathbf{I}-y\mathbf{R})^{-1}
\mathbf e_j,
\]
whose leading critical divergence is
\[
\left\langle
S_{\rm ETAS}(j)
\right\rangle
\sim
C_j
(yq)^{d_*(j)-1}
(1-y\lambda)^{-d_*(j)}.
\]
These statements concern the mean cascade and its critical divergence.
For all first-moment dynamical quantities, magnitude averaging
renormalizes the interaction matrix according to
$\mathbf R\mapsto y\mathbf R$, and hence moves the critical point from
$\lambda=1$ to $y\lambda=1$,
together with a non-universal amplitude factor. The distributional
critical behavior is not fixed by $y$: it depends on the leading
nonlinear singularity generated by the complete fertility distribution,
as derived in the next section.
}

\section{Cascade-size distribution in the ETAS model}

We now extend the previous generating-function derivation to the
magnitude-dependent ETAS process. This step exposes the separation that
is invisible at the level of the mean: the graph and its critical depth
remain structural properties, whereas the mark distribution selects the
local branching singularity that is recursively nested along the graph.

As in the finite-variance case, we label the classes along a directed
critical chain by $a=1,\ldots,d$, with $a=1$ denoting the terminal
downstream class. We define
\begin{equation}
H_a(z)
=
\sum_{S=1}^{\infty}
p_a(S)z^S,
\end{equation}
and introduce
\begin{equation}
\epsilon_a
=
1-H_a(z),
\qquad
x=1-z.
\end{equation}
For the marked ETAS process, the offspring generating function has the
non-analytic expansion
\begin{equation}
F(u)=1-u+C_\gamma u^\gamma+o(u^\gamma),
\qquad 1<\gamma<2.
\end{equation}
For the directed chain $3\rightarrow2\rightarrow1$, the offspring generating functions become
\begin{align}
F_1(H_1)
&=
1-\epsilon_1
+C_\gamma\epsilon_1^\gamma
+o(\epsilon_1^\gamma),\\
F_2(H_1,H_2)
&=
1-\epsilon_2-q\epsilon_1
+C_\gamma(\epsilon_2+q\epsilon_1)^\gamma
+o[(\epsilon_2+q\epsilon_1)^\gamma],\\
F_3(H_2,H_3)
&=
1-\epsilon_3-q\epsilon_2
+C_\gamma(\epsilon_3+q\epsilon_2)^\gamma
+o[(\epsilon_3+q\epsilon_2)^\gamma].
\end{align}
The total-tree generating functions satisfy
\begin{equation}
H_a(z)=zF_a.
\end{equation}
For type $1$,
\begin{equation}
1-\epsilon_1
=
(1-x)
\left[
1-\epsilon_1
+C_\gamma\epsilon_1^\gamma
+o(\epsilon_1^\gamma)
\right],
\end{equation}
which immediately gives
\begin{equation}
x\simeq C_\gamma\epsilon_1^\gamma.
\end{equation}
Hence
\begin{equation}
\epsilon_1
\sim
K_1x^{1/\gamma},
\qquad
K_1=C_\gamma^{-1/\gamma},
\end{equation}
recovering the scalar marked branching law
\begin{equation}
p_1(S)\sim S^{-1-\frac1\gamma}.
\end{equation}

For type $2$,
\begin{equation}
x+q\epsilon_1
\simeq
C_\gamma(\epsilon_2+q\epsilon_1)^\gamma.
\end{equation}
Near $z=1$,
\[
\epsilon_2\sim x^{1/\gamma^2},
\qquad
\epsilon_1\sim x^{1/\gamma},
\]
so that $\epsilon_2\gg\epsilon_1$. Consequently,
\begin{equation}
q\epsilon_1
\simeq
C_\gamma\epsilon_2^\gamma,
\end{equation}
yielding
\begin{equation}
\epsilon_2
\sim
K_2x^{1/\gamma^2},
\end{equation}
with
\begin{equation}
K_2
=
q^{1/\gamma}
C_\gamma^{-\left(\frac1\gamma+\frac1{\gamma^2}\right)}.
\end{equation}
Therefore,
\begin{equation}
p_2(S)
\sim
S^{-1-\frac1{\gamma^2}}.
\end{equation}

Proceeding identically for type $3$,
\begin{equation}
q\epsilon_2
\simeq
C_\gamma\epsilon_3^\gamma,
\end{equation}
which gives
\begin{equation}
\epsilon_3
=
1-H_3(z)
\sim
K_3(1-z)^{1/\gamma^3},
\end{equation}
where
\begin{equation}
K_3
=
q^{\frac1\gamma+\frac1{\gamma^2}}
C_\gamma^{-\left(
\frac1\gamma+
\frac1{\gamma^2}+
\frac1{\gamma^3}
\right)}.
\end{equation}

Applying standard singularity analysis,
\begin{equation}
p_3(S)
\sim
\frac{K_3}
{\gamma^3
\Gamma\!\left(1-\frac1{\gamma^3}\right)}
S^{-1-\frac1{\gamma^3}}~.
\end{equation}
The scalar marked-process exponent,
\begin{equation}
\tau_{\rm scalar}=1+\frac1\gamma,
\end{equation}
is therefore replaced by the non-normal exponent
\begin{equation}
\tau_{d_*=3}=1+\frac1{\gamma^3}.
\end{equation}
Since $1<\gamma<2$, one has
\begin{equation}
\frac1{\gamma^3}<\frac1\gamma,
\qquad
\tau_{d_*=3}<\tau_{\rm scalar},
\end{equation}
showing that directed non-normal amplification systematically produces
substantially heavier critical tails than the scalar marked process.
More generally, a seed with critical depth $d_*(j)$ has tail exponent
$\tau_j=1+\gamma^{-d_*(j)}$. Thus the mean dynamics depends on marks only
through $y$, while the critical cascade-size law depends on the full
fertility tail through $\gamma$; the feed-forward architecture controls
how many times that local singularity is composed.

\section{Numerical simulations}

We test the analytical predictions by direct Monte Carlo simulations of both the unmarked multitype branching process and the marked ETAS process. Each realization starts from a single ancestor of prescribed type $j_0$. For the unmarked process, a parent of type $j$ generates a Poisson-distributed number of offspring of type $i$ with mean $R_{ij}$.
For the marked ETAS process, a parent of type $j$ and magnitude $m$ generates a Poisson-distributed number of offspring of type $i$ with mean
$ R_{ij} e^{\alpha(m-m_{\min})}$.
Each offspring magnitude is independently drawn from the
Gutenberg--Richter distribution
\begin{equation}
p(m)
=
\beta e^{-\beta(m-m_{\min})},
\qquad
m\ge m_{\min}.
\end{equation}
The branching procedure is iterated generation by generation until the
cascade becomes extinct. For each realization, we record the total
cascade size $S$, including the initial ancestor.

We use the three-type matrix
\begin{equation}
\mathbf R=
\begin{pmatrix}
\lambda_1 & q & 0\\
0 & \lambda_1 & q\\
0 & 0 & \lambda
\end{pmatrix},
\label{eq:SI_matrix_simulations}
\end{equation}
where the off-diagonal elements describe a preferential propagation pathway
$
3\rightarrow2\rightarrow1.
$

We investigate two regimes. In the compensated case, we set
\[
\lambda_1=\lambda-q,
\]
so that each row has the same total weight \(\lambda\). In the homogeneous multicritical case, we set
\[
\lambda_1=\lambda.
\]
In both cases, \(q\ge0\) controls the strength of the feed-forward coupling.
In all cases \[
\rho(\mathbf R)=\lambda~.
\]
We always assume that the magnitude of the initial earthquake is set to $m_0=m_{\min}$ and indicate with $j_0$ its type. 
To test these theoretical predictions, we fix the Gutenberg--Richter exponent to $b=1$ and consider three different productivity exponents,
$
\alpha=0.77,\;0.81,\;0.83$, corresponding to $\lambda=1.19,\;0.98,\;0.90$,
chosen so that the effective branching ratio remains close to criticality (\(B_r\simeq1\)).

For each value of $\alpha$, we consider four values of the coupling parameter,
$q=0,\;0.1,\;0.2,\;0.3$,
keeping $\lambda_1=\lambda$.
For the largest coupling ($q=0.3$), we additionally investigate two reference cases:
\begin{itemize}
\item
$\lambda_1=\lambda-q$ with $j_0=3$;
\item
$\lambda_1=\lambda$ with $j_0=2$.
\end{itemize}

\begin{figure}[htbp]
    \centering
    \IfFileExists{ps_alpha077.png}
      {\includegraphics[width=\columnwidth]{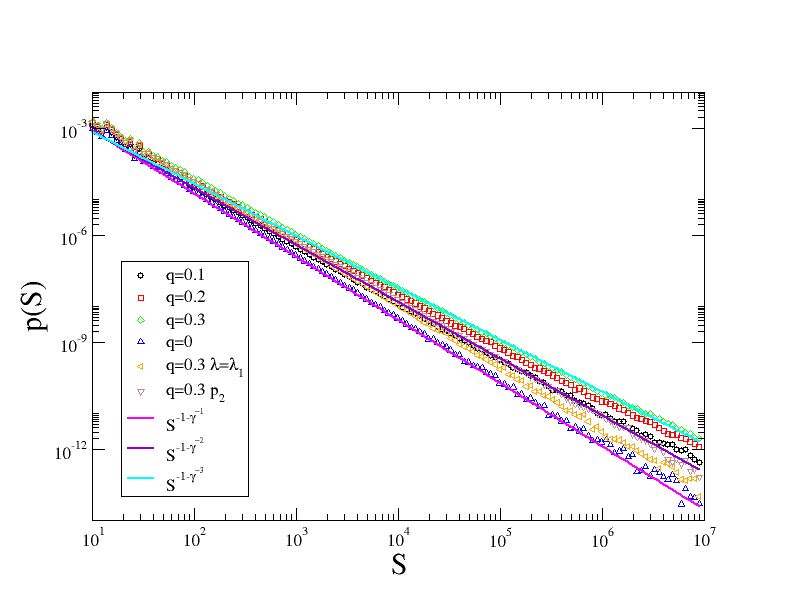}}
      {\fbox{\parbox{0.9\columnwidth}{\centering Missing simulation asset:
      \texttt{ps\_alpha077.png}}}}
    \caption{Distribution of the total cascade size $p(S,j)$ for $\alpha=0.77$ ($\lambda=1.19$). Symbols denote numerical results, while solid lines correspond to the theoretical predictions. Different curves illustrate the three asymptotic exponents $S^{-1-1/\gamma}$, $S^{-1-1/\gamma^2}$ and $S^{-1-1/\gamma^3}$ (see legend).}
    \label{figs1}
\end{figure}

\begin{figure}[htbp]
    \centering
    \IfFileExists{ps_alpha081.png}
      {\includegraphics[width=\columnwidth]{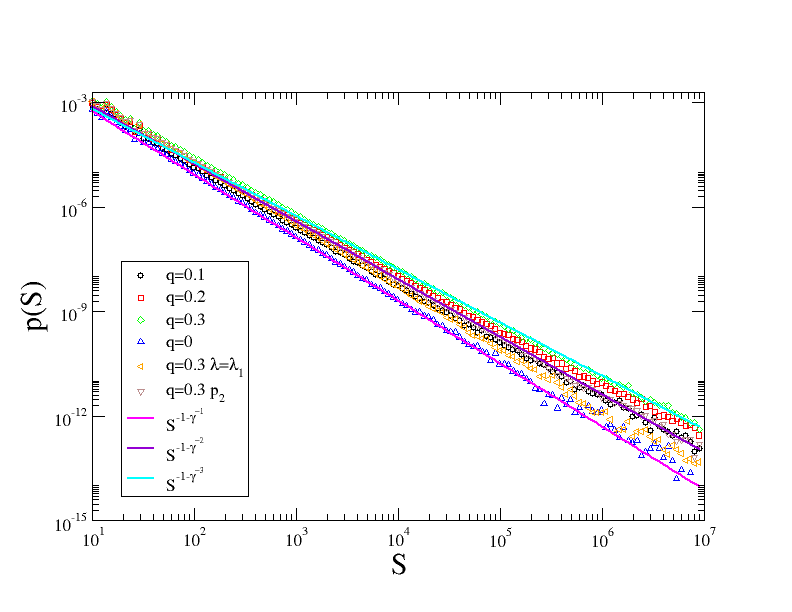}}
      {\fbox{\parbox{0.9\columnwidth}{\centering Missing simulation asset:
      \texttt{ps\_alpha081.png}}}}
    \caption{Same as Fig.~\ref{figs1} for $\alpha=0.81$ ($\lambda=0.98$).}
    \label{figs2}
\end{figure}

\begin{figure}[htbp]
    \centering
    \IfFileExists{ps_alpha083.png}
      {\includegraphics[width=\columnwidth]{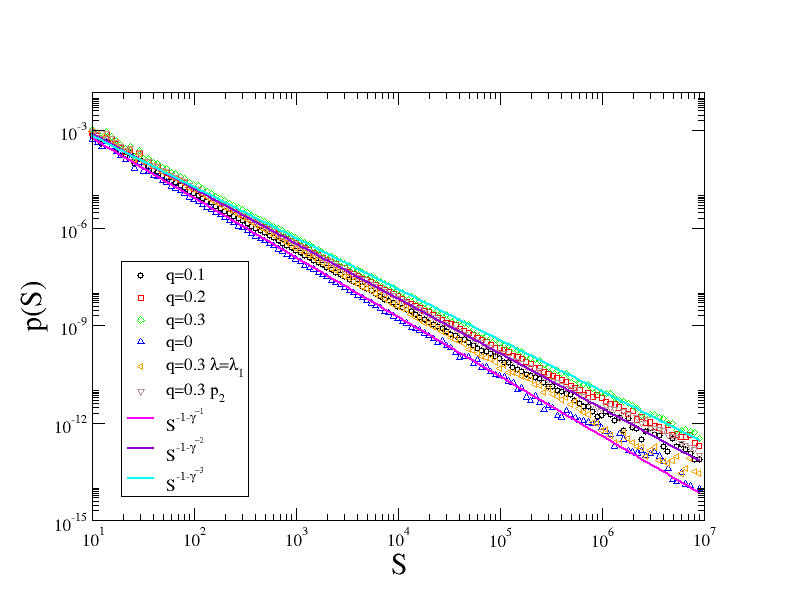}}
      {\fbox{\parbox{0.9\columnwidth}{\centering Missing simulation asset:
      \texttt{ps\_alpha083.png}}}}
    \caption{Same as Fig.~\ref{figs1} for $\alpha=0.83$ ($\lambda=0.90$).}
    \label{figs3}
\end{figure}

Figures~\ref{figs1}--\ref{figs3} compare the Monte Carlo simulations with the analytical predictions. For all three values of $\alpha$, the numerical distributions are found to be in excellent agreement with the predicted power law behavior. In particular, the simulations clearly reproduce the three different scaling exponents predicted by the theory,
\[
S^{-1-1/\gamma},
\qquad
S^{-1-1/\gamma^2},
\qquad
S^{-1-1/\gamma^3}.
\]
At the same time, we recover the standard isotropic exponent when $\lambda_1=\lambda-q$ with $p_3(S)\simeq3p_{\rm scalar}(S)$, where \(p_{\rm scalar}(S)\) denotes the corresponding scalar-process distribution.

\begin{figure}[htbp]
    \centering
    \IfFileExists{ps_alpha045.png}
      {\includegraphics[width=\columnwidth]{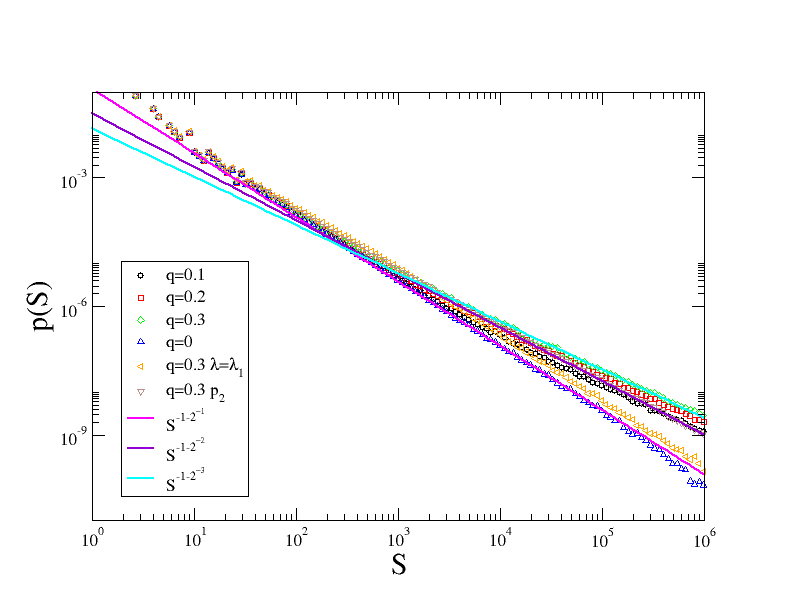}}
      {\fbox{\parbox{0.9\columnwidth}{\centering Missing simulation asset:
      \texttt{ps\_alpha045.png}}}}
    \caption{Same as Fig.~\ref{figs1} for $\alpha=0.45$  ($\lambda=2.84$).}
    \label{figs4}
\end{figure}

Fig.\ref{figs4} explores the case $\gamma>2$ and we observe that the theoretical scaling behaviors \[
S^{-1-1/2},
\qquad
S^{-1-1/4},
\qquad
S^{-1-1/8},
\] 
are recovered in numerical simulations. 

\end{document}